\documentclass[fleqn,twoside]{article}
\usepackage{espcrc2}
\usepackage{graphicx}

\newcommand{\AmS}{{\protect\the\textfont2
  A\kern-.1667em\lower.5ex\hbox{M}\kern-.125emS}}
\title{QGSJET-II: towards reliable description of very high energy 
         hadronic interactions}

\author{S.~Ostapchenko\address[MCSD]{Institut f\"ur Experimentelle Kernphysik,
University of Karlsruhe, 76021 Karlsruhe,  Germany}%
        \address{D.V. Skobeltsyn Institute of Nuclear Physics, 
         Moscow State University, 119992 Moscow, Russia}%
        \thanks{Now at Forschungszentrum Karlsruhe, Institut f\"ur Kernphysik,
76021 Karlsruhe,  Germany.}%
\thanks{ This work has been supported in part  by the German Ministry
for Education and Research (BMBF, Grant 05 CU1VK1/9).}}

\begin{document}

\begin{abstract}
Since a number of years the QGSJET model has been successfully 
used by different 
groups in the field of high energy cosmic rays. Current work is  
devoted to the first general update of the model. The key  
improvement is connected to an account for non-linear interaction  
effects which are of crucial importance for reliable model  
extrapolation into ultra-high energy domain. 
The proposed formalism allows to obtain a consistent description
of hadron-hadron cross sections and hadron structure functions
and to treat non-linear effects explicitely  
 in individual hadronic and nuclear collisions.
Other ameliorations 
concern the treatment of low mass diffraction, employment of  
realistic nuclear density profiles, and re-calibration of model  
parameters using a wider set of accelerator data. 
\vspace{1pc}
\end{abstract}

\maketitle

\section{INTRODUCTION}

All experimental studies of cosmic rays (CR) of energies above $10^{14}$
 eV are based on an indirect method: properties of the primary particles
are reconstructed on the basis of measured characteristics of 
 extensive air showers (EAS), induced by them in the atmosphere.
 The quality of such a reconstruction depends strongly 
 on the present understanding of complicated EAS
 physics, especially concerning its backbone -- hadron cascade
 process. By consequence, Monte Carlo (MC) modelization of high energy
 interactions plays an important role in CR investigations.
 
 Among other hadronic MC generators, employed in EAS simulations,
   the QGSJET model \cite{kal93,kal94,kal97} has been extensively 
  used in the field, being applied for projecting new experiments, 
  analyzing and interpreting data of various experimental installations.
  Being originally based on the Quark-Gluon String model \cite{kai82}
  picture of high energy interactions, it has been generalized to treat
  nucleus-nucleus interactions \cite{kal93} and semihard processes, using
  so-called "semihard Pomeron" approach \cite{kal94,kal97} (see also
\cite{otw97,dre99,ost02}).

Current work is devoted to another important development of the scheme - a
treatment of non-linear interaction effects. The latter are described by
so-called enhanced Pomeron diagrams \cite{gri68,gri69,kan73,car74}
and proved to be of extreme importance for a correct treatment of very
 high energy hadronic interactions. The approach proposed here is based on the
 assumption that corresponding effects are dominated by "soft" (low
 momentum transfer) partonic processes \cite{dre01,ost03}. For the first time,
essential enhanced contributions have been re-summed to all orders,
both for so-called uncut (elastic scattering) diagrams and for various
unitarity cuts of those diagrams, corresponding to particular final states
of the interaction \cite{ost05}. Based on the obtained solutions, a new 
hadronic interaction model QGSJET-II has been developed, explicitely treating 
the corresponding effects in individual hadronic (nuclear) collisions.

\section{QGSJET MODEL}

\begin{figure}[t] 
\begin{center}
  \includegraphics[width=7cm,height=2.5cm,angle=0]{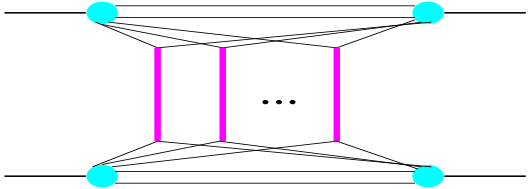} 
\end{center}
\vspace*{-1.cm}
   \caption{A general multi-Pomeron contribution to hadron-hadron scattering
amplitude; elementary scattering processes (vertical thick lines)
are described as Pomeron exchanges.\label{multiple}} 
\end{figure}

\begin{figure}[t] 
\begin{center}
  \includegraphics[width=7cm,height=2.5cm,angle=0]{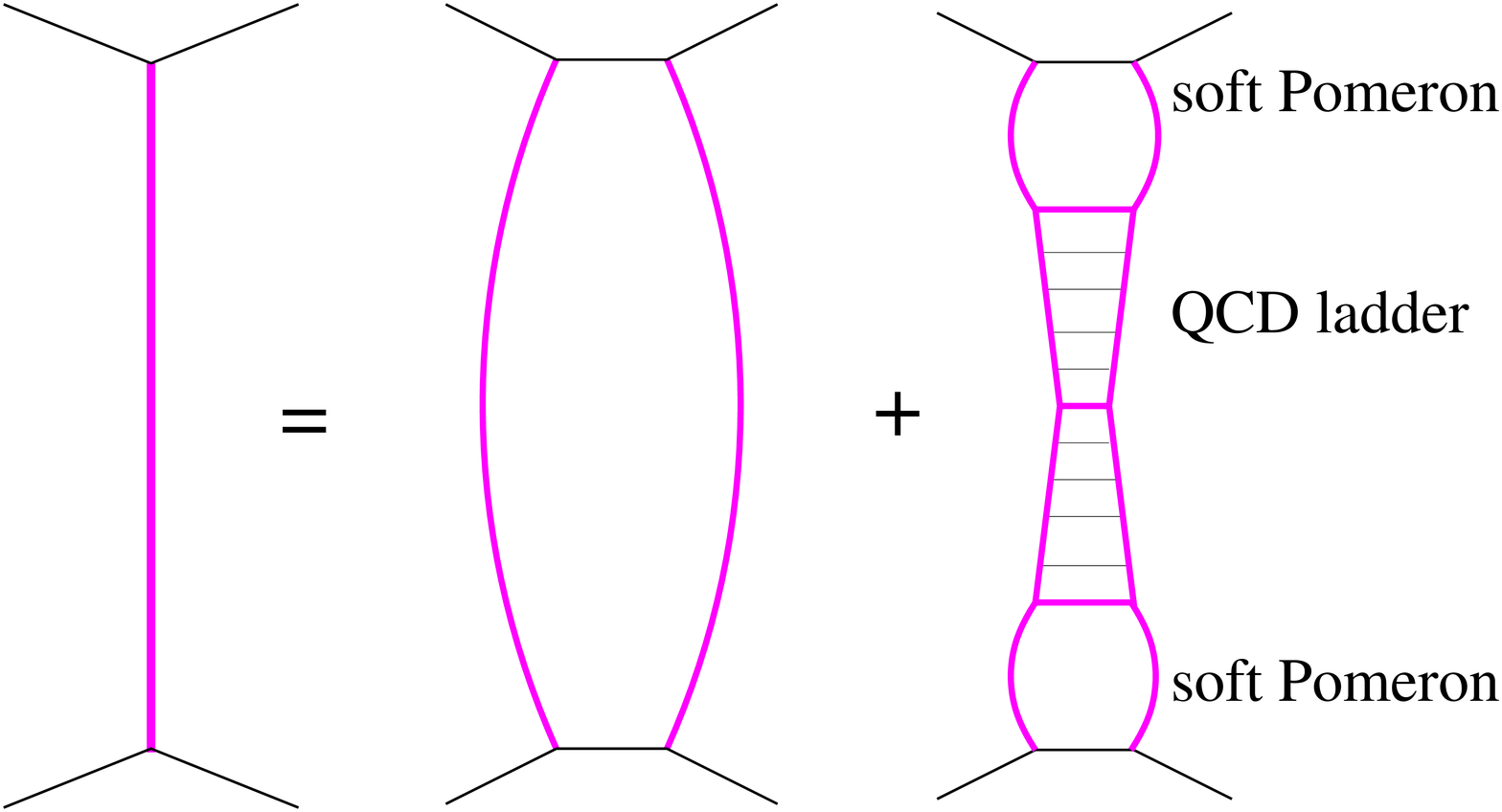} 
\end{center}
\vspace*{-1.cm}
   \caption{A "general Pomeron" (l.h.s.) consists of the "soft" 
and "semihard" Pomerons - correspondingly the 1st and the
2nd contributions on the r.h.s. \label{genpom}} 
\end{figure}

QGSJET model treats hadronic and  nuclear collisions
in the framework of Gribov's reggeon approach \cite{gri68,gri69} -- as multiple
scattering processes -- Fig. \ref{multiple}, 
where individual scattering contributions are
described phenomenologically as Pomeron exchanges. 
The latter correspond
to microscopic parton (quark and gluon) cascades, which mediate the
interaction between the projectile and the target hadron, and consist of two
parts: "soft" and "semihard" Pomerons -- Fig. \ref{genpom}. The former,
described by the corresponding 
eikonal $\chi _{\rm P}(s,b)$, represents a purely "soft"
cascade of partons of low virtualities $|q^2|<Q_0^2$, with $Q_0^2$ being some
chosen virtuality cutoff, whereas the latter, with the eikonal 
$\chi _{\rm sh}(s,b)$, corresponds to a cascade which at least partly develops
in a high virtuality ($|q^2|>Q_0^2$) region and is typically represented
by a piece of QCD ladder sandwiched between two soft Pomerons
 \cite{kal94,kal97,otw97}.

Calculating various unitarity cuts of elastic scattering diagrams of 
 Fig. \ref{multiple} according to Ab\-ram\-ovskii-Gribov-Kancheli (AGK) 
 cutting rules  \cite{agk} one can obtain expressions for total and inelastic
 cross sections as well as for relative probabilities of particular interaction
 configurations, e.g., for a given number of elementary inelastic processes
 ("cut" Pomerons), all being expressed via the total eikonal
$\chi _{\rm tot}(s,b)=\chi _{\rm P}(s,b)+\chi _{\rm sh}(s,b)$. 
 This allows in turn to perform a modelization of hadronic
(nuclear) interactions via a MC method  \cite{kal97}: starting from sampling
a particular configuration, at some impact parameter $b$ (nucleon coordinates)
and with some numbers of "cut" soft and semihard Pomerons; performing 
energy-momentum sharing between elementary inelastic interactions and between
 soft and hard partonic processes; simulating explicitely perturbative parton
 cascades; and finishing with a hadronization of strings stretched between
 constituent partons (Pomeron "ends") of the projectile and the target, and
 between final on-shell partons resulted from the QCD cascades above. 

\section{NON-LINEAR EFFECTS}
\begin{figure}[b] 
\begin{center}
  \includegraphics[width=7cm,height=2.5cm,angle=0]{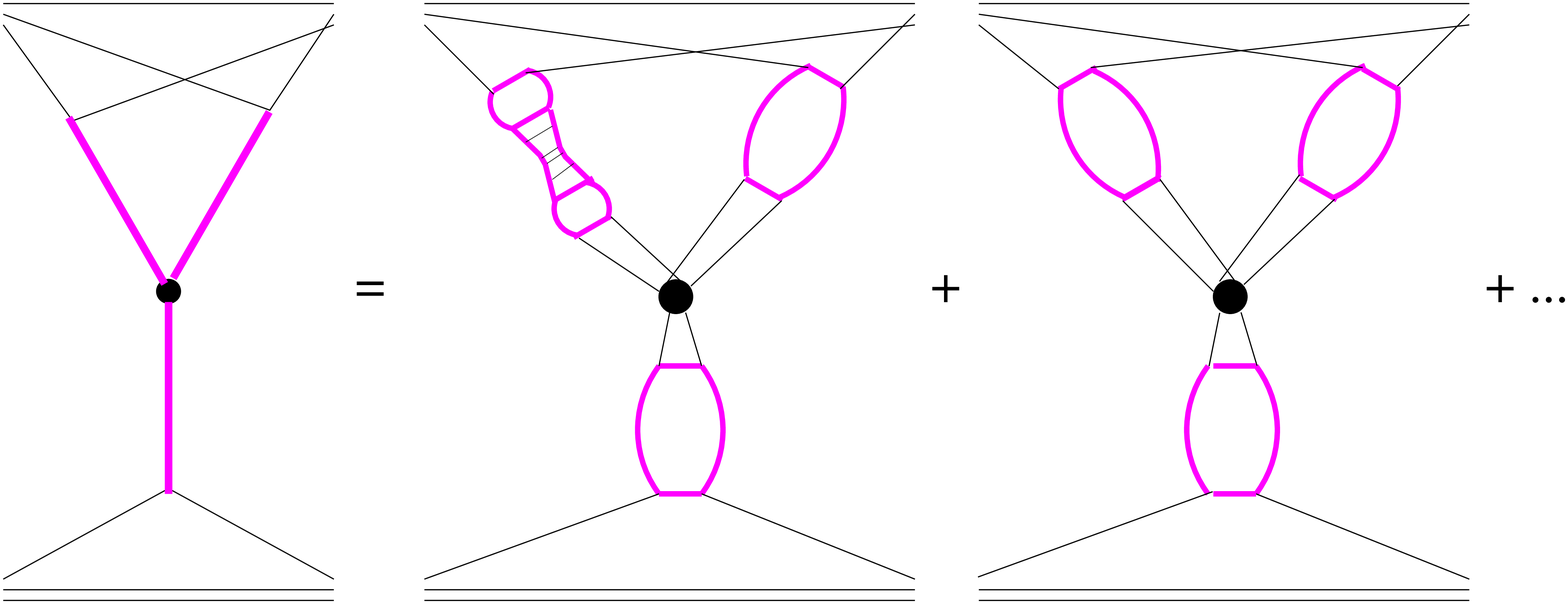} 
\end{center}
\vspace*{-1.cm}
   \caption{Contributions to the triple-Pomeron vertex from interactions between
soft and semihard Pomerons.\label{3p-vertex}} 
\end{figure}

The above-described scheme is based on the assumption that individual parton
cascades, described as Pomeron exchanges, proceed independently of each other.
This condition cease to be valid in the limit of high energies
and small impact parameters of the interaction, where a large number
of elementary scattering processes occurs and corresponding underlying
parton cascades strongly overlap and interact with each other, the 
 effects being described as Pomeron-Pomeron interactions
 \cite{gri68,gri69,kan73,car74}. We make a basic assumption
that the latter are dominated by partonic
processes at comparatively low virtuality scale, $|q^{2}|<Q_{0}^{2}$
\cite{dre01,ost03}. 

\begin{figure}[t] 
\begin{center}
  \includegraphics[width=7cm,height=2.5cm,angle=0]{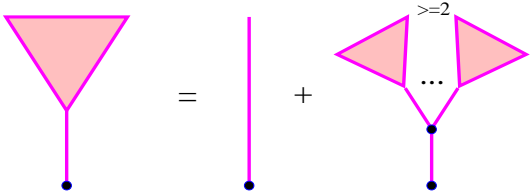} 
\end{center}
\vspace*{-1.cm}
   \caption{Recursive equation for a {}``fan'' diagram contribution 
    (constituent partons not shown). \label{ffan}} 
\end{figure} 

Then multi-Pomeron vertexes involve only interactions between soft
Pomerons or between soft "ends" of semihard Pomerons -- Fig. \ref{3p-vertex}.
At not too high energies it is sufficient to consider the 
 contributions with just one vertex of that kind.
However, with the energy increasing one has to re-sum all essential
 higher order corrections. 
 
\begin{figure}[t] 
\begin{center}
  \includegraphics[width=7cm,height=2.5cm,angle=0]{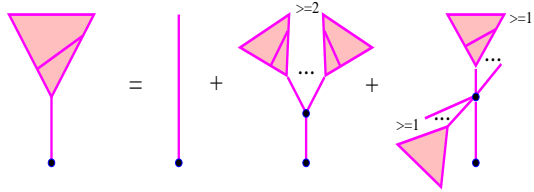} 
\end{center}
\vspace*{-1.cm}
   \caption{Recursive equation for a "generalized fan" contribution.\label{freve}} 
\end{figure}

\begin{figure*}[htb] 
\begin{center}
  \includegraphics[width=14.5cm,height=2.5cm,angle=0]{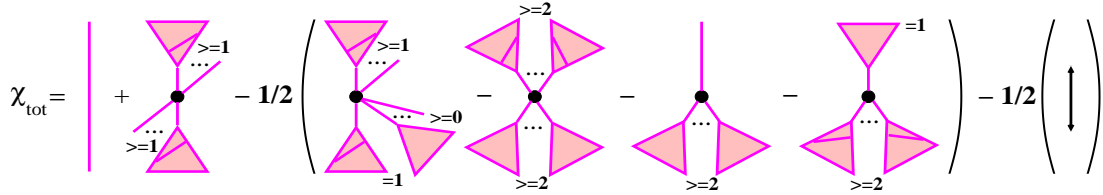} 
\end{center}
\vspace*{-1.cm}
   \caption{Total hadron-hadron interaction eikonal, including enhanced
   diagram contributions.\label{hheik}} 
\end{figure*}

 We start from calculating the "fan" diagram
 contribution, given by a recursive graphical equation of Fig. \ref{ffan}.
In addition, we define a "generalized fan" contribution as shown in 
Fig. \ref{freve}; the latter includes also vertexes with both "fans" connected
to the projectile and ones connected to the target. Then,  neglecting 
subdominant contributions of Pomeron loops, the final result for
hadron-hadron interaction eikonal can be represented in the graphical form
as shown in Fig. \ref{hheik} \cite{ost05}. Correspondingly, 
hadron-hadron elastic scattering
amplitude is now given by the sum of diagrams of Fig. \ref{multiple}, with
simple Pomeron exchanges (vertical lines in the Figure) being replaced by the
full set of diagrams of  Fig. \ref{hheik}.

To calculate probabilities of various interaction configurations one has to
consider all unitarity cuts of the diagrams of  Fig. \ref{hheik} and to re-sum
contributions of cuts of certain types.
The results obtained have been used to construct a new MC procedure
which allows to sample explicitely interaction events of complicated 
topologies
and to account for all screening corrections for particular cut Pomeron
configurations \cite{ost05}.

The treatment of hadron-nucleus and nucleus-nucleus collisions is realized
in a similar way, without introducing any additional parameters. There,
an important aspect is that different Pomerons belonging to the same 
enhanced graph
may be coupled to different nucleons of the projectile (target). 
As a consequence, non-linear screening effects are stronger in nuclear
case; in particular, this leads to a violation of the superposition picture
for nucleus-induced air showers even for average EAS characteristics
 \cite{pylos2}.

\section{DISCUSSION}

\begin{figure}[t] 
\begin{center}
  \includegraphics[width=7cm,height=3.9cm,angle=0]{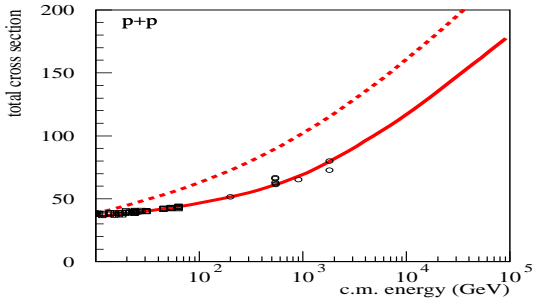} 
\end{center}
\vspace*{-1.cm}
   \caption{Total $pp$ cross section calculated with (full curve) and without
(dashed curve) enhanced diagram contributions. 
The  data are from \cite{cas98}.\label{sig-pp}} 
\end{figure}

\begin{figure}[t]
\begin{center}
  \includegraphics[width=7cm,height=3.9cm,angle=0]{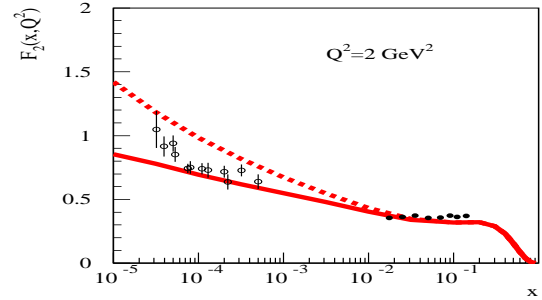} 
\end{center}
\vspace*{-1.cm}
\caption{Proton SF $F_{2}$ calculated with (full curve) and without (dashed
curve) enhanced diagram contributions.
 The data are from \cite{h1,zeus,nmc}.
\label{f2comp}}
\end{figure}

Apart from the basic development, described above, the new model included 
 realistic nuclear density parameterizations, individually
 for each nucleus \cite{kal99},  and more reliable two-component low mass
 diffraction treatment  \cite{kal93}. All model parameters
  have been re-calibrated using a wider set of accelerator data.
  The obtained results for hadron-nucleus interactions and for EAS
 characteristics will be presented elsewhere \cite{pylos2}. 
Here we illustrate the importance of Pomeron-Pomeron
 interactions  calculating total proton-proton cross section
both with and without enhanced diagram contributions --
Fig. \ref{sig-pp}.
Correspondingly,  proton structure function (SF)  $F_{2}(x,Q_0^{2})$, 
$Q_0^{2}=2$ GeV$^2$,
again calculated with and without
enhanced diagrams, 
is shown in Fig. \ref{f2comp} (charm contribution not included).%

It is important to note that the contribution of semihard processes
to the interaction eikonal can no longer be expressed in the usual factorized
form, which was assumed so far in all hadronic MC models.
The non-factorizable contributions arise from graphs
where at least one Pomeron is exchanged in parallel to the parton
hard process, with the simplest example given by the 1st diagram on
the r.h.s. in Fig. \ref{3p-vertex}. 
 At the same moment, due to the AGK
 cancellations \cite{agk} such diagrams give zero contribution
to inclusive high-$p_{t}$ jet spectra and the scheme preserves 
the QCD factorization picture for  inclusive jet production.

In conclusion, the new model is based on a consistent treatment of
 enhanced Pomeron diagrams and allows to get agreement between the 
 measured hadronic
 cross sections and structure functions and to
account for non-linear interaction effects explicitely  
 in individual hadronic and nuclear collisions.

\end{document}